# Magnetic Anisotropy and Metamagnetic Transitions in $Er_3Pt_2Sb_{4.55}$ with A Distorted Square Net Lattice

*Dylan Correll,*[a] *Chaoguo Wang,*[a] *Xin Gui* [a*]

[a] Department of Chemistry, University of Pittsburgh, Pittsburgh, PA, 15260, USA

## *Abstract*

Magnetic materials with square-net sublattices are of great interest due to its potential to realize magnetic frustration. Here we report the crystal growth and structural, magnetic, thermodynamic and electronic transport properties of a new rare-earth-based intermetallic compound, $Er_3Pt_2Sb_{4.55(5)}$. The new material possesses a distorted square-net framework of Er. Magnetic properties measurements suggest anisotropic behavior, long-range antiferromagnetic ordering and metamagnetic transitions based on a $J_{eff}$ = ½ $Er^{3+}$ motif. A magnetic structure is proposed based on the observed magnetic and thermodynamic properties. This new material expands the $R_3Pt_2Sb_{4+x}$ family with distorted square-net lattice of rare-earth elements and offers a new opportunity to study the relationship between magnetic ordering, crystal-electric field effect and crystal structure in rare-earth-based compounds.

## *Introduction*

Geometrical frustration provides an important route to intriguing quantum states including classical frustrated magnetism,[1–6] quantum spin liquid/ice[2,7–16] and superconductivity.[17–20] Rare-earth-based intermetallic compounds offer a particularly versatile chemical platform for studying geometrical-frustration-induced magnetic frustration because localized 4*f* moments are located at metallic frameworks where magnetic interactions are mediated by conduction electrons and strongly modified by crystal-electric-field anisotropy.[21,22] Representative examples include RAgGe (R = rare-earth elements) where R ions form distorted Kagome networks that support spin-ice-like behavior, complex field-induced states and topologically nontrivial features.[23–25] Moreover, square-net-like rare-earth frameworks are of interest because nearest-neighbor antiferromagnetic interactions on an ideal square lattice are not geometrically frustrated, whereas distortions, inequivalent crystal sites, further-neighbor interaction etc. can introduce competing magnetic constraints.[26–28] $R_3Pt_4Ge_6$ (R = rare-earth elements) possesses a crystal structure described as an intergrowth of $YIrGe_2$- and $BaAl_4$-type slabs with distorted square-net lattices of rare-earth elements,[29–32] which have been shown to exhibit antiferromagnetically ordered states for a selection of rare-earth members.[30,33,34] More recently, a related compound, $Dy_{3.00(1)}Pt_2Sb_{4.48(2)}$ was reported as an antimonide analogue with missing Pt layers and substantial crystal defects.[35] In this material, four crystallographically distinct Dy sites form a magnetic sublattice containing square-net-like and triangular motifs, while it also exhibits antiferromagnetic ordering near 15 K with strong magnetic anisotropy and metamagnetic transitions. However, the extensive defects on the Dy sites make further studies on its magnetic properties difficult.

Here, we report the synthesis, structure and magnetic properties of the isostructural analogue, $Er_3Pt_2Sb_{4.55(5)}$. It contains three chemically ordered Er sites while forming a slightly distorted square-net lattice, in addition to the linear chains of each Er site itself. Magnetic properties measurements suggest the existence of $J_{eff}$ = ½ on $Er^{3+}$ based on the ground-state doublet, anisotropic behavior, magnetic correlation that is formed at relatively high temperatures, and long-range antiferromagnetic ordering on two of the Er sites down to 2 K. By analyzing the magnetic, thermodynamic and electronic transport properties, we proposed a magnetic structure for $Er_3Pt_2Sb_{4.55(5)}$. Moreover, assisted by heat capacity measurements, we obtained a more quantitative and in-depth understanding of the complex magnetism in the $R_3Pt_2Sb_{4+x}$ family compared to the previous report. This work also extends the $R_3Pt_2Sb_{4+x}$ material family and highlights how local rare-earth coordination can provide anisotropic and possibly site-selective magnetic ordering in complex rare-earth intermetallics.

## *Experimental Details*

**Single Crystal Growth:** A self-flux method with excess Sb was used to obtain single crystals of $Er_3Pt_2Sb_{4.55(5)}$. Er powder (99.9%, ~40 mesh, Sigma Aldrich), Pt powder (>99.9%, ~325 mesh, Thermo Scientific) and Sb powder (99.5%, ~100 mesh, Thermo Scientific) in the molar ratio of 3:2:50 were mixed in an alumina crucible in an Ar-filled glovebox. The crucible was put into a quartz tube, with quartz wool and some glass pieces above to assist the separation of the Sb flux from the crystals after cooling. The tube was purged with argon, evacuated (<0.015 Torr) and sealed. It was heated in a furnace to 900 °C and held for 12 hours followed directly by another heat treatment at 1100 °C for two days. The furnace was then slowly cooled at the rate of 1 °C per hour to 750 °C, where it remained for a few hours then the Sb flux was centrifuged out. Water quenching the tube afterwards solidified the Sb flux in the wool plug for simple separation from the crystals. Air stable striped crystals with dimensions up to 1.07 x 0.17 x 0.05 $mm^3$ were obtained.

**Structure and Phase Determination:** The single crystal X-ray diffraction studies were carried out on a Bruker D8 Quest ECO diffractometer equipped with APEX5 software and Mo radiation ($\lambda_{K\alpha}$ = 0.71073 Å). The crystals were mounted on a Kapton loop with glycerol. The direct method and full-matrix least-squares on $F^2$ procedure within the SHELXTL package were employed to solve the crystal structure.[36,37]

**Scanning Electron Microscopy (SEM)-Energy-Dispersive X-ray Spectroscopy (EDS):** Compositional analysis was performed via scanning electron microscopy (SEM) with energy-dispersive spectroscopy (EDS) on a Zeiss Sigma 500 VP SEM instrument with Oxford Aztec X-EDS. Samples were mounted on carbon tape before loading into the SEM chamber. Multiple points and areas were examined from each sample to get the Er: Pt: Sb ratio. Samples were analyzed at 20 kV, and the spectra were collected for 120 seconds to get the chemical composition.

**Physical Property Measurement:** The Quantum Design Dynacool Physical Property (PPMS) was used to measure the DC magnetization from 2 to 300 K under a field of 0.1 T using oriented single crystal samples with the equipped ACMS II option. Field-dependent magnetization data were collected in the range −9 to 9 T under different temperatures. Temperature dependence of magnetic susceptibility was measured from 2 to 300 K under an external magnetic field with zero field cooling protocol. Heat capacity was measured using a standard relaxation method in the PPMS from 1.9 to 35 K. Oriented single crystal samples were utilized for physical properties measurement. Resistivity measurements were conducted in the PPMS from 2 to 300 K using the four-probe method, with platinum wires attached to the representative sample using silver paste.

## *Results and Discussion*

**Crystal Structure and Chemical Composition**: Single crystal X-ray diffraction was performed to determine the crystal structure of the striped crystals. The crystallographic data, including atomic positions, site occupancies, and refined anisotropic displacement parameters (and equivalent isotropic thermal displacement parameters) are listed in Table 1, and Tables S1 and S2 in Supporting Information (SI). A new material with a formula of $Er_3Pt_2Sb_{4.55(5)}$ was determined to crystallize in a monoclinic space group, $P2_1/m$ (No. 11), isostructural to the previously reported $Dy_{3.00(1)}Pt_2Sb_{4.48(2)}$.[35]

**Table 1.** Single crystal structure refinement for $Er_3Pt_2Sb_{4.55(5)}$ at 300 K.

| Refined Formula | $Er_3Pt_2Sb_{4.55(5)}$ |
|---|---|
| F.W. (g/mol) | 1445.31 |
| Space group; Z | $P2_1/m$; 2 |
| $a$ (Å) | 8.6082 (8) |
| $b$ (Å) | 4.3006 (4) |
| $c$ (Å) | 12.8978 (12) |
| $\beta$ (°) | 99.572 (3) |
| V (Å$^3$) | 470.83 (8) |
| Extinction Coefficient | 0.00023 (5) |
| θ range (°) | 2.399-31.568 |
| No. reflections; $R_{int}$ | 12828; 0.0395 |
| No. independent reflections | 1755 |
| No. parameters | 74 |
| $R_1$: $\omega R_2$ ($I$>2δ($I$)) | 0.0275; 0.0597 |
| Goodness of fit | 1.118 |
| Diffraction peak and hole (e$^-$/ Å$^3$) | 1.898; -2.542 |

The crystal structure of $Er_3Pt_2Sb_{4.55(5)}$ is shown in Figure 1a and Figure S1 in SI, consisting with three crystallographically equivalent Er sites, two Pt sites and seven Sb sites. This structure is closely related to that of a previously reported $Ln_3Pt_4Ge_6$ family (S.G. $P2_1/m$) but notably lacks two Pt layers present in the parent structure type consistent with the reduced Pt stoichiometry observed here, as described in a previous study.[35] The Pt atoms occupy the fully ordered interstitial positions between Er-Sb slabs, bridging adjacent layers. Residual electron intensity was observed during structure refinement near the Sb' site in Figure 1a and thus was modeled as three partially occupied, split-interstitial-like Sb sites marked as Sb5/Sb6/Sb7 site, as shown in Figure S1 in SI. The summed occupancy of the split sites is 55(5)%, indicating that the defective sites are partially occupied.

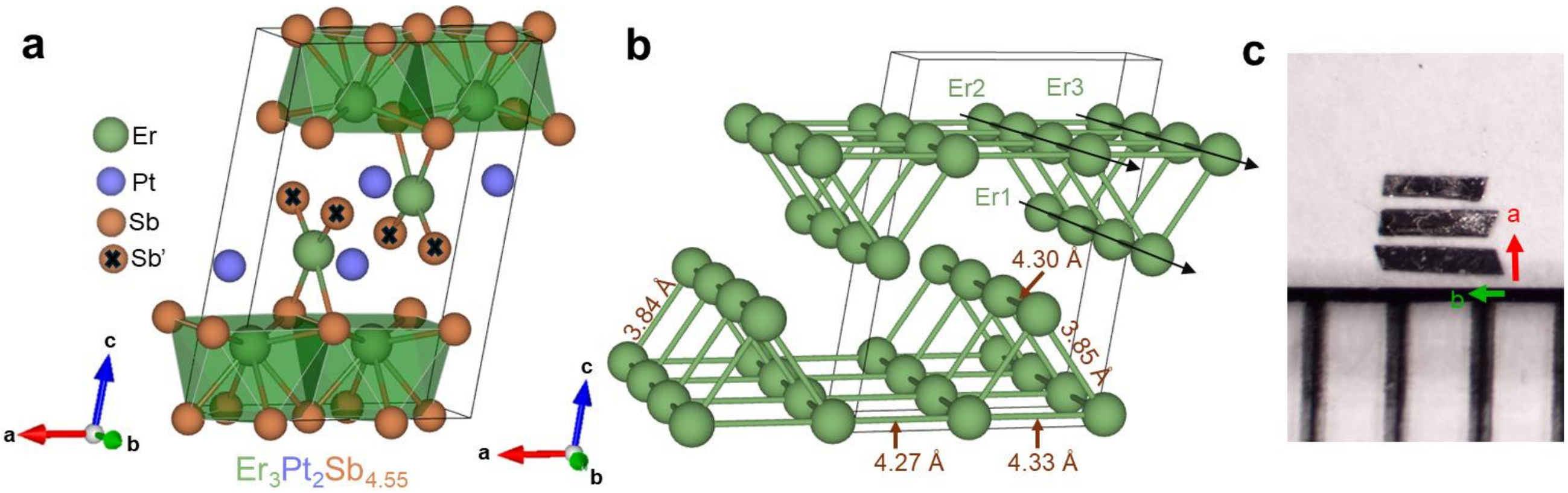

**Figure 1. a.** Crystal structure of $Er_3Pt_2Sb_{4.55(5)}$ with Er (green), Pt (blue), Sb (orange), and partially occupied Sb′ (crossed orange) sites indicated. The layered stacking of Er–Sb slabs along the c-axis is highlighted by the polyhedral representation about the Er sites. **b**. The Er sublattices, illustrating the three crystallographically distinct Er sites (Er1, Er2, Er3) and the characteristic nearest-neighbor Er–Er distances. Black arrows indicate the directions of the Er1/Er2/Er3 linear chains along the *b* axis. (c) Optical image of a few representative crystals of $Er_3Pt_2Sb_{4.55(5)}$, with the crystallographic *a* (red arrow) and *b* (green arrow) directions indicated.

The coordination types of three Er sites can be seen in Figure 1a. Er1 forms a $Er@Sb_4$ distorted square planar geometry, while Er2 and Er3 possess a similar eight-coordinated geometry by constructing $Er@Sb_8$ square antiprisms. Figure 1b presents the Er framework, in which the three crystallographically distinct Er sites form linear chains along the *b* axis with an intrachain Er-Er distance of ~4.30 Å. The Er2 and Er3 sites further construct a slightly distorted square-net lattice, with alternating interchain Er-Er distance of ~4.27 Å and ~4.33 Å stacked along the *a* axis. The shortest Er-Er distances are observed between Er1 and Er2/Er3 chains (~3.85 Å for Er1-Er2 and ~3.84 Å for Er1-Er3). A picture of the striped crystals is shown in Figure 1c. The longer edge of the crystals was determined to be the *b* axis while the shorter edge was the *a* axis, indicating that the crystals prefer to grow along the direction of Er chains.

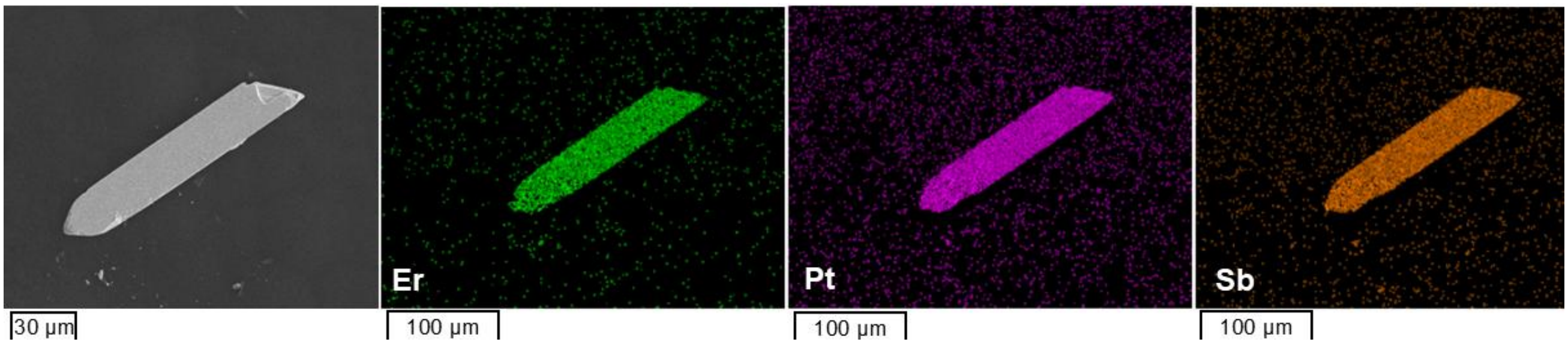

**Figure 2.** SEM image of $Er_3Pt_2Sb_{4.55(5)}$ and elemental EDS maps.

To confirm the chemical composition of the synthesized crystals, SEM-EDS was performed on representative single crystals. An SEM image of a representative crystal is shown in Figure 2. Elemental mapping of Er, Pt, and Sb reveals a uniform distribution of all three elements across the crystal, with no evidence of any secondary phases or concentration gradients at the resolution of the EDS measurement. The homogeneous distribution of each element is consistent with a single-phase crystalline material and supports the stoichiometry determined from single crystal X-ray diffraction refinement. Quantitative compositions were extracted from point EDS measurements averaged over multiple crystals and can be found in Table S3. Because Er, Pt and Sb are all heavy elements with large atomic numbers and appreciably different X-ray scattering factors, the site assignments and refined occupancies from single-crystal X-ray diffraction are expected to be robust. Here, we will use the formula determined by X-ray diffraction, $Er_3Pt_2Sb_{4.55(5)}$, in the following context and data analyses.

**Magnetic Anisotropy, Antiferromagnetic Ordering and Spin Reorientation**: Magnetic property measurements were performed on $Er_3Pt_2Sb_{4.55}$ single crystals with the magnetic field applied along three directions: parallel to the *a* axis (H//*a*), parallel to the *b* axis (H//*b*), and perpendicular to the *ab* plane (H$\perp ab$). The orientation of representative stripe-like crystals is shown in Figure 1c. The temperature dependence of magnetic susceptibility ($\chi$) and inverse $\chi$ with an applied magnetic field of 1000 Oe from 2 to 300 K are shown in Figure 3. All curves were collected under zero field cooling (ZFC) conditions. A modified Curie-Weiss (CW) law

$$\chi = \chi_0 + \frac{C}{T - \theta_{CW}}$$

was applied to fit the $\chi$(T) curves where $\chi_0$ and C are independent of temperature (the latter related to the effective moment and the former to the core diamagnetism and temperature independent paramagnetic contributions such as Pauli paramagnetism), and $\theta_{CW}$ is the CW temperature. $\chi_0$ was then subtracted to obtain the temperature-dependent Curie-Weiss component of magnetic susceptibility, while it is 0.00125 emu/Oe/mol, -0.00496 emu/Oe/mol and -0.07205 emu/Oe/mol for H//*a*, H//*b* and H⊥*ab*.

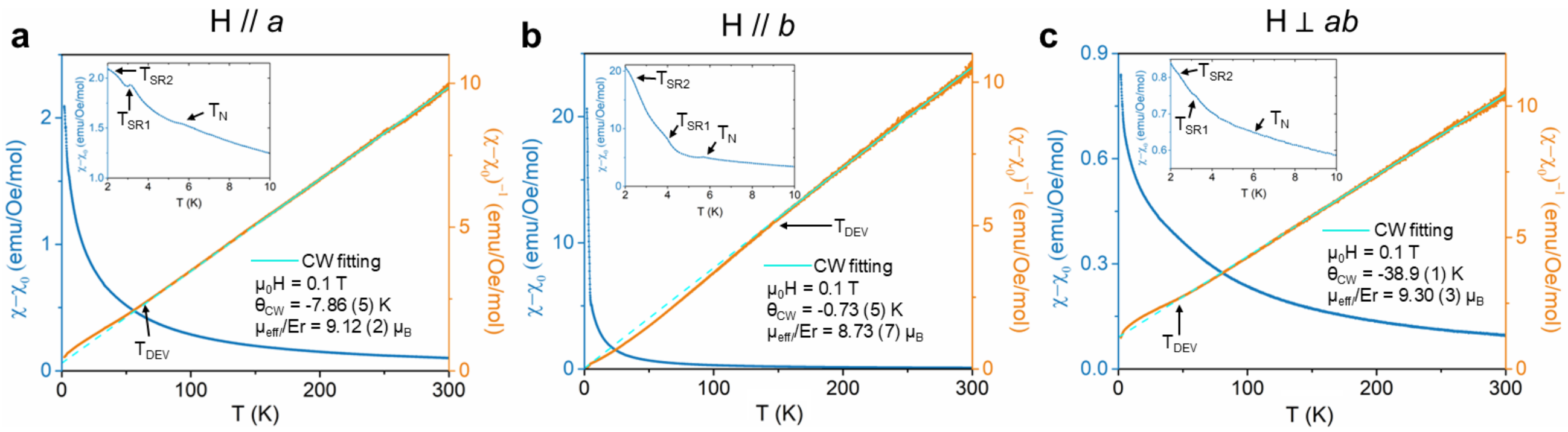

**Figure 3.** The temperature dependence of magnetic susceptibility ($\chi$) and inverse magnetic susceptibility ($\chi^{-1}$) after subtracting $\chi_0$ when the external magnetic field is **(a)** parallel to the *a* axis, **(b)** parallel to the *b* axis, and **(c)** perpendicular to the *ab* plane. The solid cyan lines represent Curie-Weiss fitting while the dashed one is the extension. Insets show the low-temperature susceptibility below 10 K, with the long-range antiferromagnetic ordering temperature $T_N$ and the two spin reorientation temperatures $T_{SR1}$ and $T_{SR2}$ identified.

As can be seen in Figure 3, typical paramagnetic-like behavior is observed at high temperatures (> ~150 K) for all three field directions in the $(\chi - \chi_0)^{-1}$ curves. However, deviations from the linear CW behavior occur at $T_{DEV}$ ~ 60 K for H//*a*, ~ 150 K for H//*b* and ~ 50 K for H⊥*ab*, indicating that weak magnetic correlations start to form below those temperatures. The linear portion of the curves was fitted using the CW law above. The fitted $\theta_{CW}$ are -7.86 (5) K, -0.73 (5) K and -38.9 (1) K. The negative signs indicate antiferromagnetic spin-spin interaction in all three directions. The distinct values suggest strong magnetic anisotropy, indicating that the net exchange interaction in the fitted region is antiferromagnetic for H//*a* and H⊥*ab* by average, while the ferromagnetic and antiferromagnetic coupling are nearly cancelled out when H//*b*. Moreover, the effective moment ($\mu_{eff}$/Er) can be obtained by $\mu_{eff}$/Er = $\sqrt{8C/n}$ $\mu_B$, where n is the number of Er per formula unit. The fitted $\mu_{eff}$/Er was found to be 9.12 (2) $\mu_B$, 8.73 (7) $\mu_B$ and 9.30 (3) $\mu_B$, close to what is expected for $Er^{3+}$ (9.58 $\mu_B$). The potential cause for the slightly lower values compared to standard $\mu_{eff}$ of $Er^{3+}$ can be that the fitting temperature range (150 K – 300 K) may not be high enough to thermally populate all the energy levels due to the crystal electric field (CEF) effect.

Magnetic anisotropy is also confirmed by the low-temperature behaviors. As shown in Figures 3a – 3c, a sharp upturn is observed in $(\chi - \chi_0)$ when H//*b*, while a less prominent increase is seen when H//*a*. When H⊥*ab*, the low-temperature magnetic susceptibility is the smallest. These observations are consistent with the fact that *b* axis is the easy axis, while *a* is the intermediate axis and the direction perpendicular to *ab* plane is the hard axis. Note that the out-of-plane direction is not the *c* axis in $Er_3Pt_2Sb_{4.55(5)}$ with a monoclinic unit cell. Additionally, three anomalies were observed at low temperatures for all directions. A peak can be clearly seen when H//*a* and H//*b* at $T_N \sim 5.7$ K, indicating an antiferromagnetic ordering transition, consistent with the negative CW temperatures. Such a transition, however, is hardly visible when H⊥*ab*, suggesting the near absence of the formation of antiparallelly aligned spins along that direction. Two other bumps were present at lower temperatures, represented by $T_{SR1}$ and $T_{SR2}$, standing for spin-reorientation temperatures. $T_{SR1}$ and $T_{SR2}$ both emerge at slightly higher temperatures when H//*b*, compared with the other directions, consistent with the fact that *b* axis is the easy axis. For example, $T_{SR1}$ is ~3.85 K for H//*b*, ~3.16 K for H//*a* and ~3.15 K for H⊥*ab*, while $T_{SR2}$ is ~2.41 K for H//*b*, ~2.31 K for H//*a* and ~2.31 K for H⊥*ab*.

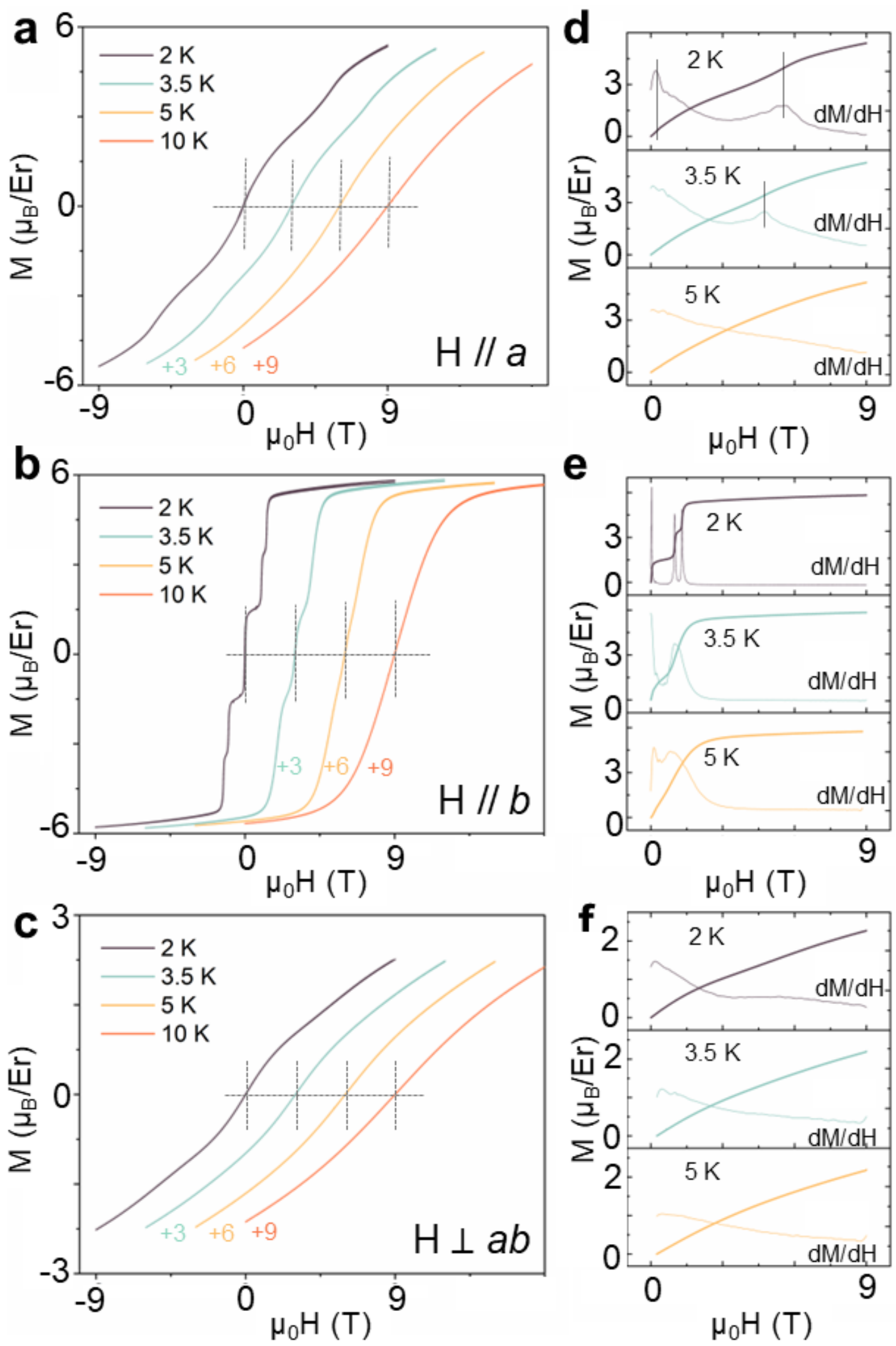

**Figure 4.** Field-dependent magnetization for when the external magnetic field is **(a)** parallel to the *a* axis, **(b)** parallel to the *b* axis, and **(c)** perpendicular to the *ab* plane. The loops measured under different temperatures have been offset by clarity. The right panels **(d–f)** show the corresponding first-derivative (dM/dH) curves at 2, 3.5, and 5 K for **(d)** parallel to the *a* axis, **(e)** parallel to the *b* axis, and **(f)** perpendicular to the *ab* plane respectively, with peaks identifying the metamagnetic transitions.

**Metamagnetic Transitions:** To understand the magnetic behaviors of $Er_3Pt_2Sb_{4.55}$ further, the field-dependent magnetization was measured under different temperatures with different field directions. Clear anisotropic behavior can be seen. As shown in Figures 4a – 4c, metamagnetic transitions were observed in $Er_3Pt_2Sb_{4.55(5)}$. Details of the transitions are illustrated in Figures 4d – 4f. At 2 K, sharp transitions were found when H//*b* at ~0.95 T and ~1.3 T, while only a broad transition was seen at ~5.5 T for H//*a* and an even broader one at ~5.6 T for H⊥*ab*. When temperature increases to 3.5 K, all metamagnetic transitions were either completed suppressed or observed under lower magnetic fields with a broader feature. For example, the one transition occurred when H//*a* can be found at ~4.7 T at 3.5 K and further became invisible at 5 K, while the one in H⊥*ab* disappeared at 3.5 K. Moreover, only one broad metamagnetic transition for H//*b* can be seen at ~1.03 T and ~0.93 T at 3.5 K and 5 K.

As can be seen in Figure 4b, the magnetization is nearly saturated at $\mu_{sat}$ ~ 5.78 $\mu_B$/Er when H//*b*, which is ~64%, i.e., 2/3, of the expected saturated moment, $\mu_{exp}$ ~ 9.0 $\mu_B$/Er, for free $Er^{3+}$ ions. Similarly, although not showing a saturation plateau yet, the magnetization at 9 T, $\mu_{9T}$, for H//*a* is ~5.37 $\mu_B$/Er, close to the value for H//*b*. However, a much smaller $\mu_{9T}$ ~ 2.26 $\mu_B$/Er was found for H⊥*ab*, supporting the observation in Figure 3 that *b* axis is the easy axis while the direction perpendicular to *ab* plane is the hard axis.

Additionally, a small coercive field (~230 Oe) exists for H//*b*, while the value is nearly negligible for the other directions. This may indicate that the ground-state magnetic structure possesses a small ferromagnetic component along the *b* axis evidence of this small coercive field can be found in Figure S2. We will further discuss this later.

**Heat Capacity**: In order to obtain more in-depth information about the magnetic ground state of $Er_3Pt_2Sb_{4.55(5)}$, heat capacity ($C_p$) measurement was conducted at zero magnetic field. As shown in Figure 5a, temperature-dependent $C_p$ was collected from 1.9 K to 35 K. Two well-resolved λ anomalies are observed at low temperature around ~5.6 and ~5.4 K, consistent with the long-range antiferromagnetic ordering at $T_N$ ~5.7 K. When further cooled down, $C_p$ (T) exhibits two small bumps at ~3.1 K and ~2.4 K, coinciding with short range ordering. The fact that the peak at $T_N$ contains two nearly equally significant peaks implies that two magnetic ordering may occur at $T_N$, which will be discussed later.

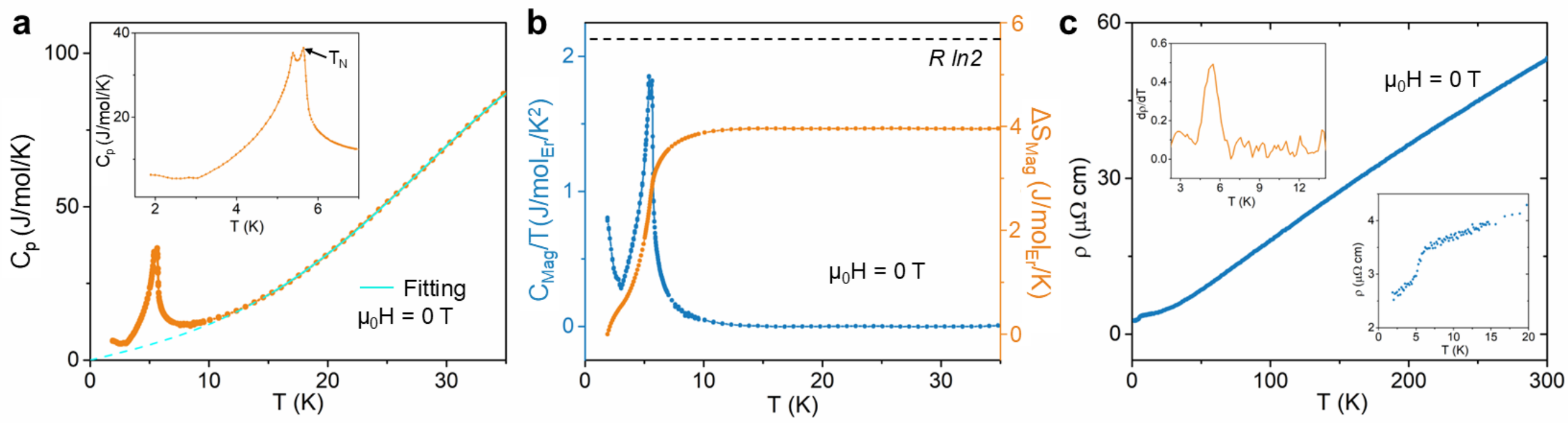

**Figure 5. (a) (Main Panel)** Heat capacity ($C_p$) of $Er_3Pt_2Sb_{4.55(5)}$ measured from 2 to 35 K under no external magnetic field with fitting shown as the cyan solid line with the dashed line as extension. **(Inset)** Enlarged view of the $C_p$ vs T curve from 2 to 7 K. **(b)** Magnonic contribution to the heat capacity ($C_{mag}$/T) and the corresponding entropy change ($\Delta S_{mag}$/Er) from 2 to 35 K where dotted line shows the Rln2 value. **(c) (Main panel)** Temperature-dependent electrical resistivity (ρ) with no magnetic field applied. **(Inset top left)** First derivative of the main panel at low temperature. **(Inset bottom right)** Enlarged view of the low-temperature region, 2 to 20 K, of ρ(T).

The heat capacity of a metal commonly contains three components: electronic, phononic, and magnonic contributions, thus, $C_p = C_{el} + C_{ph} + C_{mag}$. By subtracting $C_{el}$ and $C_{ph}$ from $C_p$, the magnonic contribution can be obtained and will provide helpful information to the magnetic quantum states. Here, we employ the following equation to fit the paramagnetic region above $T_N$,

$$C_p = \gamma T + \beta_1 T^3 + \beta_2 T^5$$

where γ is Sommerfeld coefficient, indicating electronic contribution, while β1 and β2 demonstrate phononic contributions. The best results occurred when fitting from 15 K to 35 K, which gave $\gamma$ = 0.946 (5) J/mol/K$^2$, $\beta_1$ = 0.00230 (1) J/mol/K$^4$ and $\beta_2$ = -0.00000085 (1) J/mol/K$^6$. The obtained γ is a very large value. However, because the fitted temperature range likely contains substantial crystal-electric-field effect and magnetic correlations, this value should not be interpreted as a true Sommerfeld coefficient.

After subtracting the fitted electronic and phononic background, the intrinsic magnonic contribution, $C_{mag}$/T, is shown in Figure 5b. The magnetic entropy ($\Delta S_{mag}$) can then be obtained using $\Delta S_{mag} = \int \Delta C_{mag}/T dT$, which has also been plotted in Figure 5b. As can be seen, a minor $\Delta S_{mag}$/Er ~0.5 J/mol$_{Er}$/K exists below ~3.1 K, likely corresponding to short range ordering. The major $\Delta S_{mag}$/Er contribution (~3.5 J/mol$_{Er}$/K) appears above ~3.1 K and saturates at ~10 K, originating from the long-range antiferromagnetic ordering at $T_N$. It is noteworthy that the major $\Delta S_{mag}$/Er ~ 3.5 J/mol$_{Er}$/K is only ~ 61% of Rln2, which is the expected magnetic entropy change for a *S or* $J_{eff}$ = ½ particle. Here, $Er^{3+}$ exists in $Er_3Pt_2Sb_{4.55(5)}$, as determined by magnetic properties measurements. $Er^{3+}$, although is a $S$ = 3/2 and $J$ = 15/2 particle as a free ion, as a Kramers ion, each Er site has a CEF ground-state doublet in zero field. If this doublet is well isolated from the first excited CEF level, i.e., at low temperatures, the low-temperature magnetism can be modeled using an effective $J$ = ½.[38,39] Thus, it is appropriate to compare $\Delta S_{mag}$/Er with Rln2. The fact that it is ~61% of the expected value coincides with what was observed from magnetization measurements, i.e., $\mu_{sat}$ ~ 5.78 $\mu_B$/Er when H//*b*, which is ~64%, i.e., ~2/3, of the expected saturated moment, $\mu_{exp}$ ~ 9.0 $\mu_B$/Er.

**Electronic Transport Properties**: The temperature-dependent electrical resistivity, ρ(T), of $Er_3Pt_2Sb_{4.55}$ was measured at zero field as shown in Figure 5c. ρ(T) exhibits metallic behavior over the full measured range, which decreases monotonically with decreasing temperature from about ~53 μΩ cm at 300K towards ~2.5 μΩ cm at 2 K. The residual-resistivity ratio (RRR) is ~21 for $Er_3Pt_2Sb_{4.55}$, suggesting good crystal quality. A small kink was observed at ~ 6 K, which is likely due to the long-range antiferromagnetic ordering transition. Note that a linear relation is seen above $T_N$ ~ 6 K till ~20 K, indicating that the electronic transport properties are strongly influenced by magnetic correlations or critical spin fluctuations, i.e., due to the existence of short-range magnetic correlation, consistent with the observations from χ(T) results.

**Discussion of Potential Magnetic Structure:** Given the fact that both $\mu_{sat}$ and $\Delta S_{mag}$/Er are consistent with nearly 2/3 of the Er sites are fully magnetically ordered above 2 K, it is plausible that two of the three Er sites in $Er_3Pt_2Sb_{4.55(5)}$ actually participate in the long-range antiferromagnetic ordering, supported by the double-peak behavior in $C_p$(T) at $T_N$. Moreover, Er2 and Er3 both possess an 8-coordinated square antiprism geometry, while Er1 has a distinct 4-coordinated geometry, it is reasonable to speculate that the magnetic correlation strengths within Er2 and Er3 linear chains are similar, which gives rise to a similar ordering temperature, consistent with the double peaks in $C_p$(T). Moreover, the spin direction of the magnetic ground state can be indicated by that the *b* axis is the easy axis where spins are most easily polarized by magnetic field, while no obvious $T_N$ and only a very weak spin polarization can be observed for H⊥*ab*. This leads to the possibility for two different magnetic structure possibilities. In the first, the spin directions are likely within the *ab* plane while the net moment projects mostly on the *b* axis. The observed spin reorientations thus suggest that the spins are not fully aligned along the *b* axis, i.e., there is a spin canting predominantly within the *ab* plane, which is further proved by the small coercivity observed when H//*b*. The second possibility is that the spins orient themselves antiferromagnetically parallel to the *b* axis. With the three metamagnetic transitions seen when H//*b* in Figure 4b, it can be suspected that they are from three successive transitions from the ground state with 3 spins up and 3 spins down for the first transition, 4 spins up and 2 spins down for the second transition, then to 5 spins up and 1 spin down for the third transition, and finally to a saturated ferromagnetic state with all spins aligned. The proposed magnetic structures are shown in Figure S3 in SI. However, more detailed magnetic structure will require further determination by neutron diffraction.

## *Conclusions*

In this paper, we synthesized millimeter-sized crystals of a new rare-earth-based intermetallic, $Er_3Pt_2Sb_{4.55(5)}$. Based on magnetic properties measurements, we observed an antiferromagnetic ordering transition followed by two spin-reorientation transitions. The field-dependent magnetization further revealed several metamagnetic transitions in all crystallographic directions, indicating the sensitivity of the magnetic ground state to the external magnetic field. The heat capacity measurements helped us to resolve the possible magnetic structure of $Er_3Pt_2Sb_{4.55(5)}$ where two of the three Er sites are antiferromagnetically ordered above 2 K with a $J_{eff} = ½$, while the predominant spin direction in the magnetically ordered state is along the crystallographic *b* axis. This work introduces a new member in the $R_3Pt_2Sb_{4+x}$ family with a distorted square-net framework and offers a new platform to study their complex magnetism without extensive crystal defects on the rare-earth-element sites.

## *Associated Content*

Supporting Information

The Supporting Information is available free of charge at XXX.

Atomic coordinates and anisotropic thermal displacement parameters. Crystal structure with all partially occupied Sb sites shown. The elemental analysis based on EDS. Proposed magnetic structure.

## *Author Information*

Corresponding Author: xig75@pitt.edu

Notes: The authors declare no competing financial interest.

## *Acknowledgements*

This research is supported by the startup funds from the University of Pittsburgh. SEM-EDS performed in the University of Pittsburgh Nanofabrication and Characterization Core Facility (RRID:SCR_05124) and services and instruments used in this project were graciously supported, in part, by the University of Pittsburgh.

***Supplementary Information***

# Magnetic Anisotropy and Metamagnetic Transition in $Er_3Pt_2Sb_{4.55}$ with A Distorted Square Net Lattice

*Dylan Correll,*[a] *Chaoguo Wang,*[a] *Xin Gui* [a*]

[a] Department of Chemistry, University of Pittsburgh, Pittsburgh, PA, 15260, USA

**Table of Contents**

**Table S1.** Atomic coordinates and equivalent isotropic displacement parameters for $Er_3Pt_2Sb_{4.55(5)}$ at 300 K. ($U_{eq}$ is defined as one-third of the trace of the orthogonalized $U_{ij}$ tensor (Å$^2$))

| Atom | Wyck. | Occ. | *x* | *y* | *z* | $U_{eq}$ |
|---|---|---|---|---|---|---|
| Pt1 | 2*e* | 1 | 0.46572(4) | 1/4 | 0.38033(3) | 0.00664(9) |
| Pt2 | 2*e* | 1 | 0.97461(4) | 1/4 | 0.38056(3) | 0.00665(9) |
| Er1 | 2*e* | 1 | 0.26532(5) | 1/4 | 0.56065(4) | 0.00767(10) |
| Er2 | 2*e* | 1 | 0.42062(5) | 3/4 | 0.18902(3) | 0.00834(10) |
| Er3 | 2*e* | 1 | 0.92430(5) | 3/4 | 0.18983(3) | 0.00833(10) |
| Sb1 | 2*e* | 1 | 0.18671(7) | 1/4 | 0.24717(5) | 0.00803(13) |
| Sb2 | 2*e* | 1 | 0.31142(7) | 3/4 | 0.74668(5) | 0.00703(13) |
| Sb3 | 2*e* | 1 | 0.12402(7) | 3/4 | 0.99751(5) | 0.00880(13) |
| Sb4 | 2*e* | 1 | 0.37514(7) | 1/4 | 0.00175(5) | 0.00866(13) |
| Sb5 | 2*e* | 0.259(4) | 0.6166(3) | 1/4 | 0.5624(2) | 0.0134(8) |
| Sb6 | 2*e* | 0.258(4) | 0.9146(3) | 1/4 | 0.5623(2) | 0.0128(8) |
| Sb7 | 2*e* | 0.028(4) | 0.765(3) | 1/4 | 0.559(2) | 0.014(8) |

**Table S2.** Anisotropic thermal displacement parameters for $Er_3Pt_2Sb_{4.55(5)}$.

| Atom | $U_{11}$ | $U_{22}$ | $U_{33}$ | $U_{23}$* | $U_{13}$* | $U_{12}$* |
|---|---|---|---|---|---|---|
| Pt1 | 0.0051(2) | 0.0081(2) | 0.0068(2) | 0 | 0.0017(1) | 0 |
| Pt2 | 0.0053(2) | 0.0080(2) | 0.0068(2) | 0 | 0.0018(1) | 0 |
| Er1 | 0.0059(2) | 0.0088(2) | 0.0085(2) | 0 | 0.0020(1) | 0 |
| Er2 | 0.0081(2) | 0.0089(2) | 0.0081(2) | 0 | 0.0018(2) | 0 |
| Er3 | 0.0085(2) | 0.0089(2) | 0.0079(2) | 0 | 0.0023(2) | 0 |
| Sb1 | 0.0057(3) | 0.0101(3) | 0.0087(3) | 0 | 0.0021(2) | 0 |
| Sb2 | 0.0051(3) | 0.0085(3) | 0.0078(3) | 0 | 0.0019(2) | 0 |
| Sb3 | 0.0084(3) | 0.0098(3) | 0.0086(3) | 0 | 0.0023(2) | 0 |
| Sb4 | 0.0083(3) | 0.0093(3) | 0.0086(3) | 0 | 0.0021(2) | 0 |
| Sb5 | 0.0123(13) | 0.0159(14) | 0.0128(13) | 0 | 0.0047(9) | 0 |
| Sb6 | 0.0122(13) | 0.0132(14) | 0.0133(13) | 0 | 0.0030(9) | 0 |

*For an explanation of the anisotropic thermal displacement parameters, see *The International Tables for Crystallography*[1], A. Authier editor, second edition, volume D, pages 231 to 245, John Wiley and Sons, 2014.

**Table S3.** The atomic ratios from Energy-Dispersive X-ray Spectroscopy (EDS) result of two different crystals.

| | **Er%** | **Pt%** | **Sb%** |
|---|---|---|---|
| **Number of points** | **Sample#1** | | |
| **1** | 32.55 | 22.23 | 45.22 |
| **2** | 32.39 | 22.40 | 45.21 |
| **3** | 32.57 | 22.34 | 45.09 |
| **4** | 32.48 | 22.34 | 45.18 |
| **5** | 32.62 | 22.36 | 45.02 |
| **Number of points** | **Sample#2** | | |
| **1** | 31.89 | 22.63 | 45.47 |
| **2** | 31.98 | 22.56 | 45.46 |
| **3** | 32.18 | 22.18 | 45.63 |
| **4** | 32.52 | 22.33 | 45.15 |
| **5** | 32.31 | 22.49 | 45.20 |
| **Average** | 32.35 (8) | 22.39 (4) | 45.26 (6) |
| **Normalized** | 3.00 (1) | 2.08 (1) | 4.20 (1) |

**Figure S1.** Crystal structure of $Er_3Pt_2Sb_{4.55(5)}$ with Er (green), Pt (blue), Sb (orange), and partially occupied Sb5/Sb6/Sb7 sites labeled.

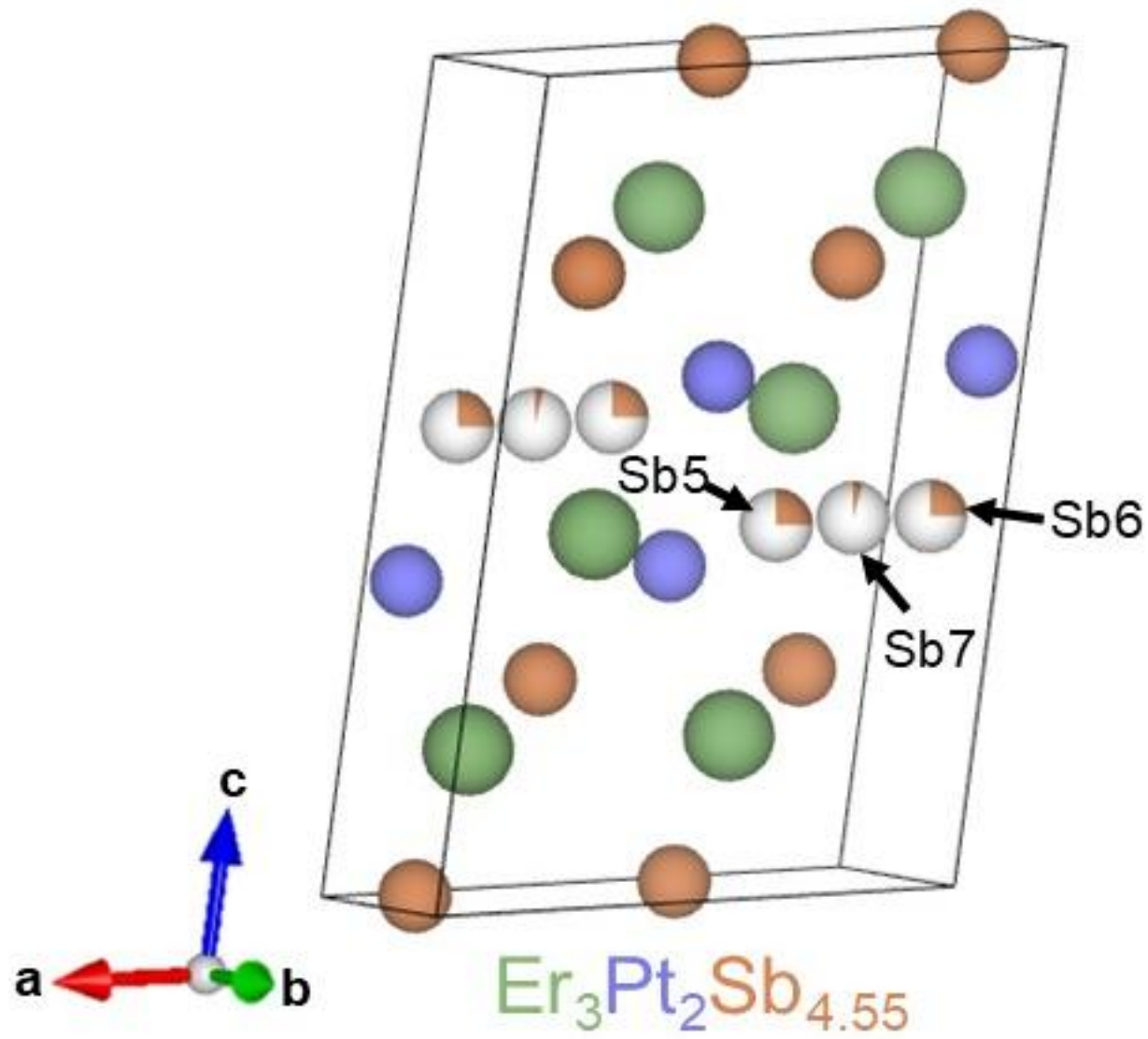

**Figure S2.** Enlarged view for the 2 K H//*b* magnetization vs field plot.

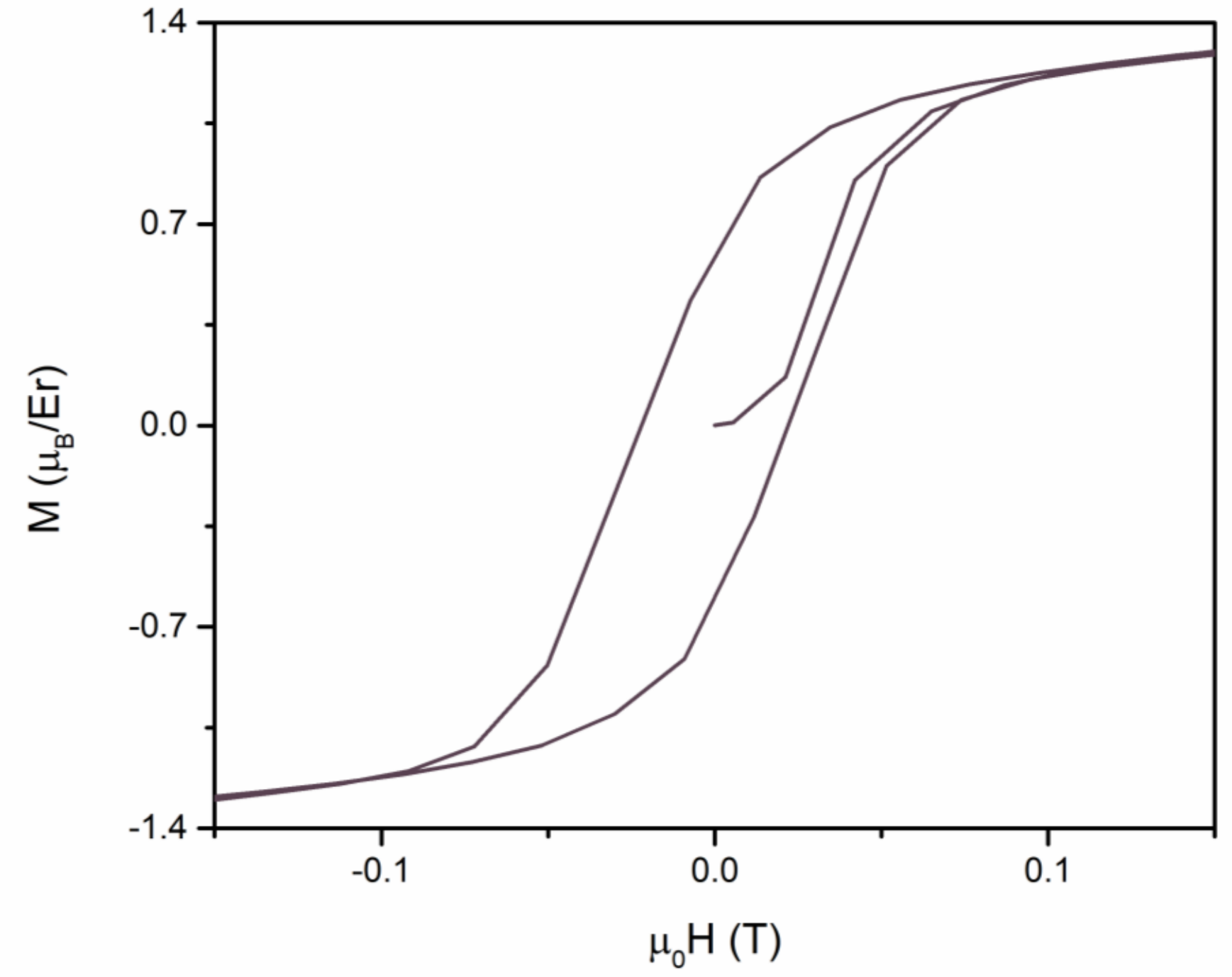

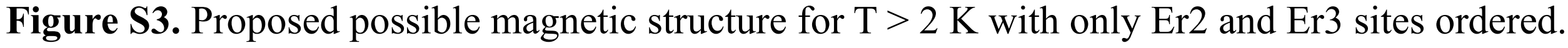

**Figure S3.** Proposed possible magnetic structure for T > 2 K with only Er2 and Er3 sites ordered.

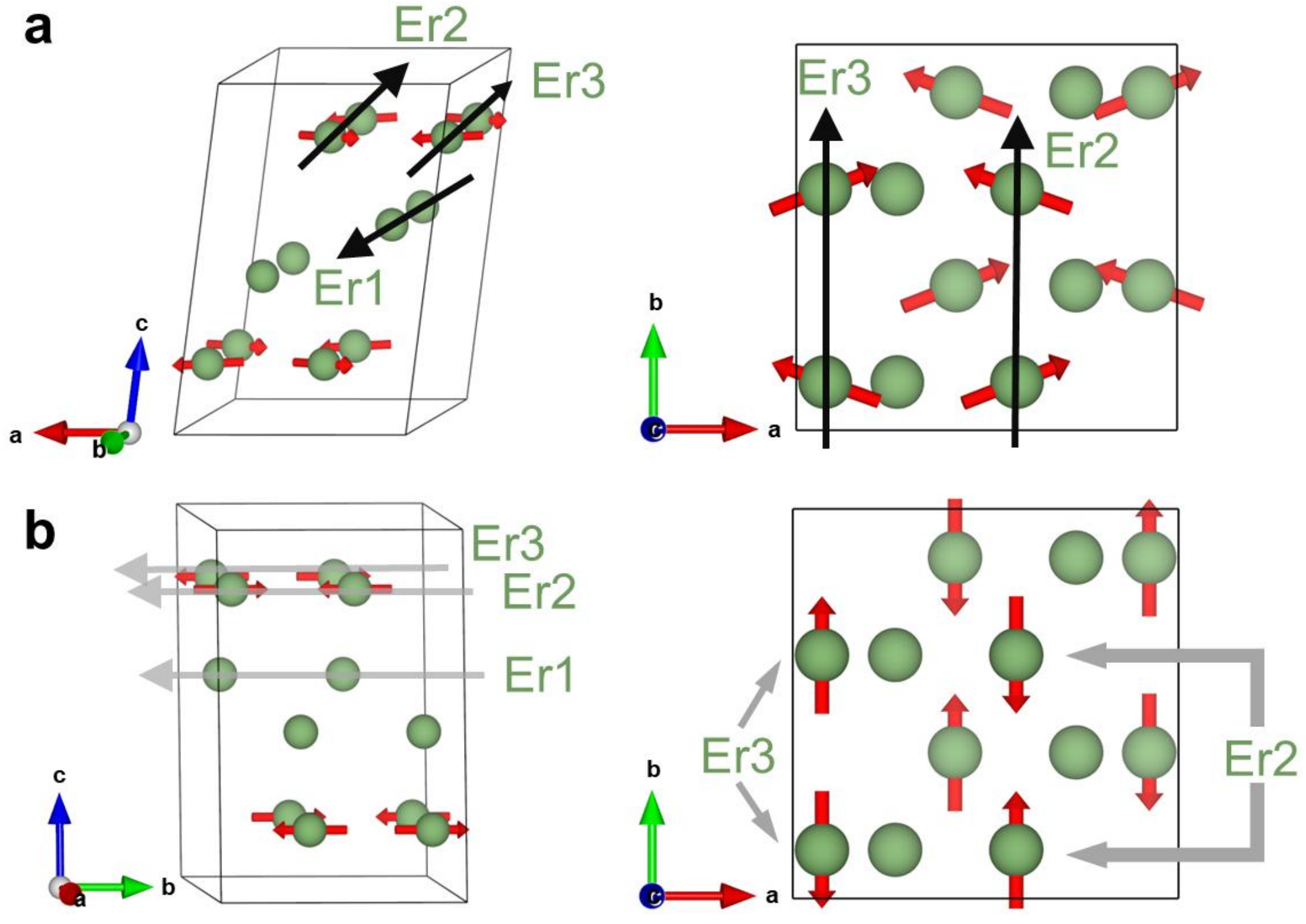